\def\year{2022}\relax
\documentclass[letterpaper]{article} 
\usepackage[strings]{underscore}
\usepackage{aaai25}
\usepackage{times}  
\usepackage{helvet}  
\usepackage{courier}  
\usepackage[hyphens]{url}  
\usepackage{graphicx} 
\usepackage[mode=match]{siunitx}
\usepackage{natbib}  
\usepackage{caption} 
\DeclareCaptionStyle{ruled}{labelfont=normalfont,labelsep=colon,strut=off} 
\usepackage{algorithm}
\usepackage[noend]{algpseudocode}
\makeatletter
\def\BState{\State\hskip-\ALG@thistlm}

\usepackage{tabularx}
\usepackage{booktabs}
\usepackage{multirow}
\usepackage{amsmath}
\usepackage{comment}
\usepackage{xcolor}
\usepackage{xspace}

\newcommand{\fullname}{RepoGenReflex\xspace}
\usepackage{newfloat}
\usepackage{listings}
\floatstyle{ruled}
\newfloat{listing}{tb}{lst}{}
\floatname{listing}{Listing}
\title{RepoGenReflex: Enhancing Repository-Level Code Completion with Verbal Reinforcement and Retrieval-Augmented Generation}
\author{
    Jicheng Wang, Yifeng He, Hao Chen
}
\affiliations{
    University of California, Davis\\

   \{jicwang, yfhe, chen\}@ucdavis.edu
}

\begin{document}

\maketitle

\begin{abstract}
In real-world software engineering tasks, solving a problem often requires understanding and modifying multiple functions, classes, and files across a large codebase. 
Therefore, on the repository level, it is crucial to extract the relevant information to achieve accurate code completion effectively. 
Existing code completion tools have achieved some success, but they struggle to optimize the retrieval and generation process dynamically. 
In this paper, we propose \fullname, a generic, dynamic, effective framework to address this challenge. 
By leveraging the Retrieval-Augmented Generation (RAG) enhanced with Verbal Reinforcement Learning (VRL), it can dynamically choose the optimal results for repository-level code completion.
\fullname uses Reflector to give directional feedback to the next loop.
\fullname chooses the optimal results stored in the Experience cache based on the RAG-VRL loop.
To validate the framework's generalization ability, we propose a new benchmark RepoGenEval, which consists of the latest, high-quality real-world repositories in line completion scenarios.
Our experiments demonstrate that \fullname achieves significant improvements after optimizing the Reflector component, resulting in enhanced accuracy and relevance of code completions. Additionally, \fullname consistently demonstrates superior performance and effectiveness across standard code completion tasks, highlighting the robustness and adaptability of our framework.
Our source code and benchmark will be publicly available.
\end{abstract}
\section{Introduction}
 
Code completion is a popular feature in modern software development. 
By automatically suggesting relevant code snippets and structures to the developer, code completion tools reduce the inclusion of irrelevant code snippets and improve the efficiency of the programmer.
However, code completion tools have fundamental limitations.
First, rule-based code completion like tools in Integrated Development Environments is difficult to develop and maintain,
and cannot be generalized to settings other than development, for example, 
writing new features or fixing bugs~\cite{robbes2008program,bruch2009learning,proksch2015intelligent}.
Other approaches like automatic patch finding by genetic programming~\cite{weimer2009automatically} extend code completion to fixing bugs. These methods depend on static analysis on the Abstract Syntax Tree~\cite{mccarthy1960recursive} and heuristic rules to guide the evolution process, but they may struggle with varying-length completions and may not dynamically adapt to new coding patterns efficiently.

To overcome these limitations,
various work~\cite{izadi2022codefill,guo2023longcoder,petryshyn2024optimizing} purposes to utilize large language models (LLMs) to generate patches for code completion.

Recent advancements in machine learning(ML) and natural language processing(NLP), especially with the development of LLMs, have led to the development of sophisticated code completion frameworks~\cite{svyatkovskiy2019pythia}.
Given an incomplete code snippet as the prompt, LLMs can generate the following tokens as completion, taking advantage of the massive code dataset it memorized at the training stage~\cite{nijkamp2022codegen,li2022competition,phan2021cotext}.
Although simply prompting LLMs with all the related context information for code completion is direct and relatively efficient,
this approach is limited by LLMs' context window size.
To generate the response successfully,
we cannot give all the related context information to LLMs in prompts.
In real-world software engineering, developers need to be aware of other files within the repository since there exists interrelated dependencies, shared utilities, configurations, and cross-API invocations resulting from modularization. Additionally, LLMs fine-tuned on labeled data excel in controlled evaluation settings but face challenges generalizing to unseen repositories without further fine-tuning. This is particularly due to the models' reliance on localized knowledge, which is often insufficient when dealing with complex, modularized codebases that span multiple files and components. These limitations highlight the need for a more adaptive and context-aware approach to repository-level code completion. Therefore, direct code completion via LLM generation is far from practical usage.

To overcome the limitations of context length and localized knowledge in reading long documents and adapting to new tasks, \citeauthor{lewis2020retrieval} proposed Retrieval-Augmented Generation (RAG), which uses retrieval to draw on relevant snippets from longer text or documents.

RAG generates contextually relevant and accurate text by retrieving relevant snippets and combining them with the prompt provided to the LLM.
RAG has been adopted in various recent attempts in LLM-based applications~\cite{cai2022recent,jimenez2023swe,sumers2023cognitive}.
RepoCoder~\cite{zhang2023repocoder} utilizes RAG to bridge the gap between static knowledge in pre-trained models and the dynamic, context-sensitive needs of real-world coding scenarios.
RepoCoder works by iteratively retrieving relevant code snippets from the repository, using them to generate accurate and contextually appropriate code. However, a major challenge remains in managing the iterative process to dynamically find out optimal results for each task.
RepoCoder relies on fixed iteration limits rather than dynamically adjusting based on the generated outputs, which can impact both accuracy and efficiency in large-scale applications.


Reinforcement learning (RL) is promising in addressing the challenge of dynamic optimization in iterative processes.
Recent research has explored the potential of RL for optimizing iterative processes, particularly through policy gradient methods~\cite{lu2022dynamic}. 
While this approach has shown state-of-the-art(SOTA) performance in tasks such as mathematical reasoning, it requires fine-tuning and updating model weights, which is both computationally expensive and limits generalization across different LLMs. 
Verbal reinforcement learning (VRL)~\cite{shinn2024reflexion}, on the other hand, provides a promising alternative by leveraging linguistic feedback to refine models without modifying their weights. VRL enables efficient optimization across various tasks including decision-making, coding, and language reasoning. 
VRL addresses the computational expense and limited generalization across different LLMs by refining models through feedback rather than fine-tuning, thus maintaining model efficiency and adaptability.

\begin{figure}[h]
    \centering
    \includegraphics[width=0.47\textwidth]{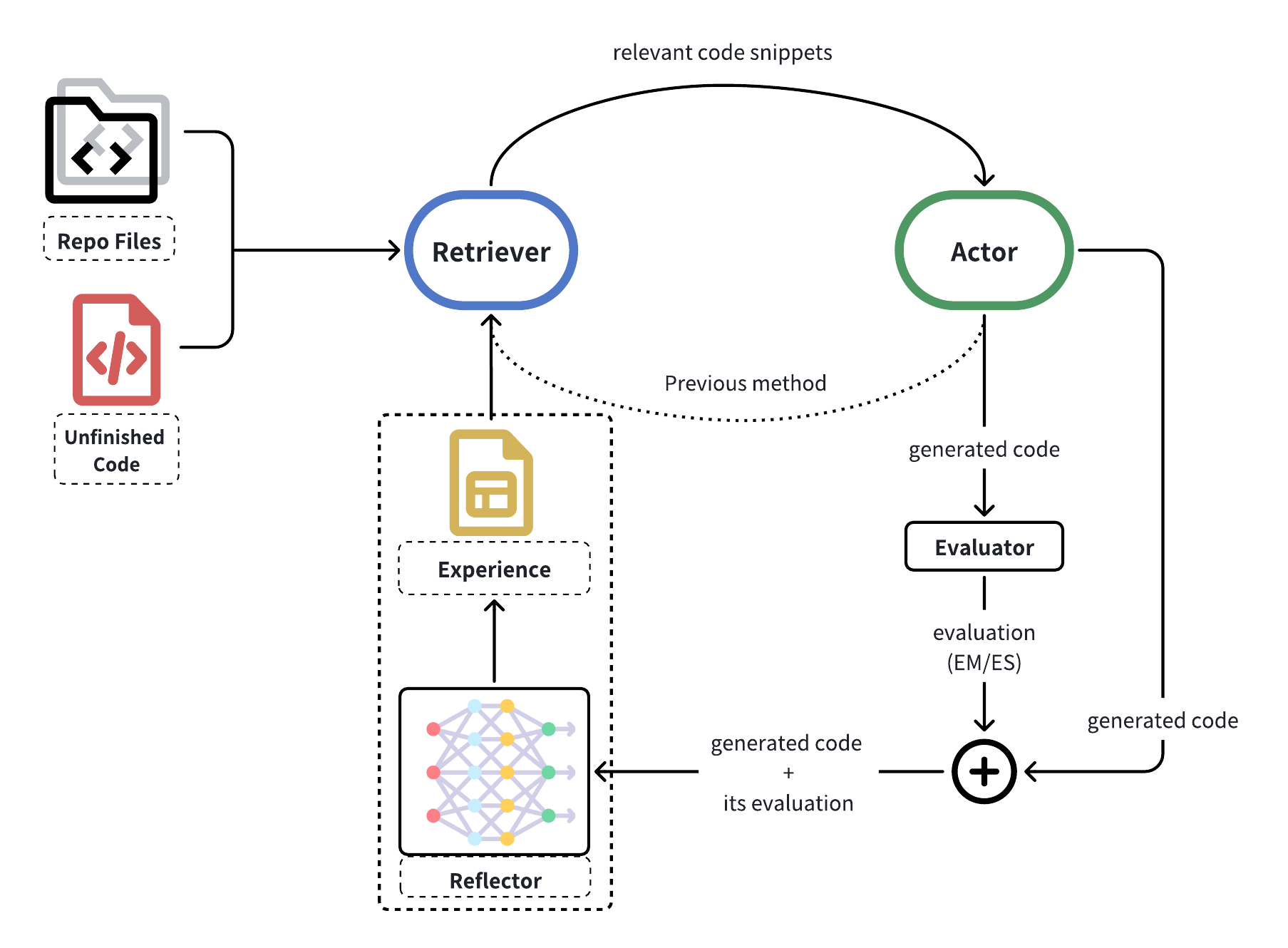}
    \caption{\small An overview of the model design, illustrating the iterative process of Retriever, Actor (Generative LLM), Evaluator, Reflector, and Experience components.}
    \label{fig:model-design}
\end{figure}

In this paper, we introduce \fullname, a novel approach that enhances the RAG framework with VRL to further refine LLM-based repository-level code completion,
optimizing retrieval and generation iterations through iterative feedback.
As demonstrated in Figure~\ref{fig:model-design}, \fullname iteratively retrieves relevant code snippets, generates possible code completion, evaluates the results by Evaluator, and gives detailed feedback using Reflector.
Reflector provides linguistic feedback, guiding the next iteration to improve the quality of the generated code. 
Each feedback is stored in the Experience component, which helps refine future iterations, ensuring that the framework continuously improves its performance without the need for model weight updates, enabling efficient optimization across various tasks.

Existing benchmarks like RepoEval~\cite{zhang2023repocoder} fall short in comprehensively evaluating the generalization ability of our framework, especially in diverse and complex real-world scenarios. RepoEval's broad selection criteria and basic evaluation methods may not fully capture the challenges of modern codebases. To address these limitations, we propose RepoGenEval, a new benchmark with stricter selection criteria for high-quality repositories. It provides a thorough evaluation of our framework's generalization and effectiveness capabilities.

Our experiments demonstrate that, by integrating VRL within the RAG framework, \fullname achieves significant improvements in optimizing the Reflector component, resulting in enhanced accuracy and relevance of code completions. Using EM and ES metrics as evaluation metrics, \fullname consistently demonstrates superior practical performance across standard code completion tasks. These results highlight the robustness and adaptability of our framework, suggesting a promising new direction for intelligent code generation tools. Our contributions can be summarized as follows:
\begin{itemize}
    \item We propose \fullname, a novel iterative retrieval-generation framework for repository-level code completion, leveraging VRL to dynamically optimize the retrieval and generation processes.
    \item We introduce an adaptive iterative process that dynamically evaluates and optimizes each iteration's result without model weight updates, ensuring efficient and continuous improvement in code completion performance.
    \item We demonstrate that \fullname surpasses traditional in-file code completion methods and outperforms current SOTA models by achieving significantly higher accuracy and relevance, establishing a robust foundation for the future development of code generation tools.

\end{itemize}
\section{Related Work}

\paragraph{Retrieval-Augmented Generation (RAG)}
RAG is a widely adopted strategy that enhances LLMs by integrating retrieval mechanisms with generative processes~\cite {lewis2020retrieval}. Traditional RAG approaches typically involve three phases: \textbf{Retrieval}, which extracts relevant information from external sources related to the task.
\textbf{Augmentation}, which enriches the input with the retrieved data to enhance context.
\textbf{Generation}, which uses the enriched input to generate a more accurate, context-aware response. RAG has been successfully applied in various NLP tasks, including question answering~\cite{siriwardhana2023improving,mao2020generation}, document summarization~\cite{parvez2021retrieval,liu2020retrieval}, and code completion~\cite{parvez2021retrieval,zhang2023repocoder,wang2024rlcoder,wu2024repoformer}. 
\paragraph{Verbal Reinforcement Learning (VRL)}
Reinforcement Learning (RL) has demonstrated its efficacy in optimizing iterative processes by dynamically adjusting the model's actions based on feedback. Traditional RL methods, such as Policy Gradient~\cite{lu2022dynamic}, require fine-tuning of model weights, which is computationally expensive and can limit the model's generalization across different tasks. VRL offers an alternative by utilizing linguistic feedback to guide the iterative refinement of models without altering their underlying weights~\cite{shinn2024reflexion}. VRL has been successfully applied in tasks such as sequential decision-making, language reasoning, and code generation. For example, the Reflexion framework~\cite{shinn2024reflexion} employs VRL to iteratively refine the outputs of LLMs, significantly improving performance without the need for weight updates. This makes VRL an attractive approach for scenarios where computational efficiency and adaptability are critical.
\paragraph{Repository-Level Code Completion}
Repository-level code completion involves generating code by considerin g the entire codebase of repositories, rather than just isolated files or functions. This approach has become a focal point for research in the code completion area and many studies have attempted to improve it.
A study by \citeauthor{shrivastava2023repository} proposes a framework that learns to generate example-specific prompts by prompt proposals. This highlights the ongoing challenges in effectively leveraging LLMs for code completion.
RepoCoder~\cite{zhang2023repocoder} leverages RAG to iteratively retrieve and utilize relevant code snippets from repositories, enhancing the accuracy and context-awareness of code completions. However, RepoCoder has a weakness in that its performance relies on fixed iteration limits, which can result in suboptimal performance when the iteration number is not dynamically adjusted based on the generated outputs' quality.

In this work, we extend the application of VRL to an RAG framework tailored for repository-level code completion. By incorporating VRL into the iterative retrieval and generation process, our approach aims to dynamically optimize each iteration based on feedback, enhancing the generalization and effectiveness of the framework without requiring model fine-tuning. This integration addresses the limitations of previous methods and provides a more flexible and powerful tool for intelligent code generation.

\section{Framework Design}

Figure \ref{fig:model-design} depicts our framework, which consists of a Retriever, Actor (Generative LLM), Evaluator, Reflector, and Experience. Together they work iteratively to generate high-quality code.


\subsection{Components}
\subsubsection{Retriever}

The Retriever component is responsible for extracting relevant code snippets from the repository files, which provide contextual information for the unfinished code. This step is crucial for grounding the language model's generation in a relevant context, thereby improving the coherence and relevance of the generated code. 

\subsubsection{Actor (Generative LLM)}

The Actor component utilizes a pre-trained language model (LM) to generate code by prompting. The initial prompt for LM consists of the retrieved snippets concatenated with the unfinished code, providing the LM with comprehensive context information.

\subsubsection{Evaluator}

The generated code is then passed to the Evaluator component, which assesses the code using Exact Match (EM) and Edit Similarity (ES) metrics. These metrics have been widely adopted in code completion area~\cite{lu2021codexglue,zhang2023repocoder}. 
The EM metric is a binary measure that takes the value of 1 if the predicted code exactly matches the ground truth code and 0 otherwise. This provides a straightforward assessment of whether the model's output is correct. To capture more subtle differences between the generated and ground truth code to provide a more fine-grained evaluation,
we employ the ES metric,
\[
ES = 1 - \frac{\operatorname{Lev}(\hat{Y}, Y)}{\max\left( |\hat{Y}|, |Y| \right)}
\]

where \(\text{Lev}\) denotes the Levenshtein distance~\cite{levenshtein1966binary}, a metric that measures the minimum number of single-character edits (insertions, deletions, or substitutions) required to change one string into another. This formula ensures that the ES score considers both the size of the predicted and ground truth code, providing a more nuanced evaluation of generation quality.


\begin{figure}[h]
    \centering
    \includegraphics[width=0.48\textwidth]{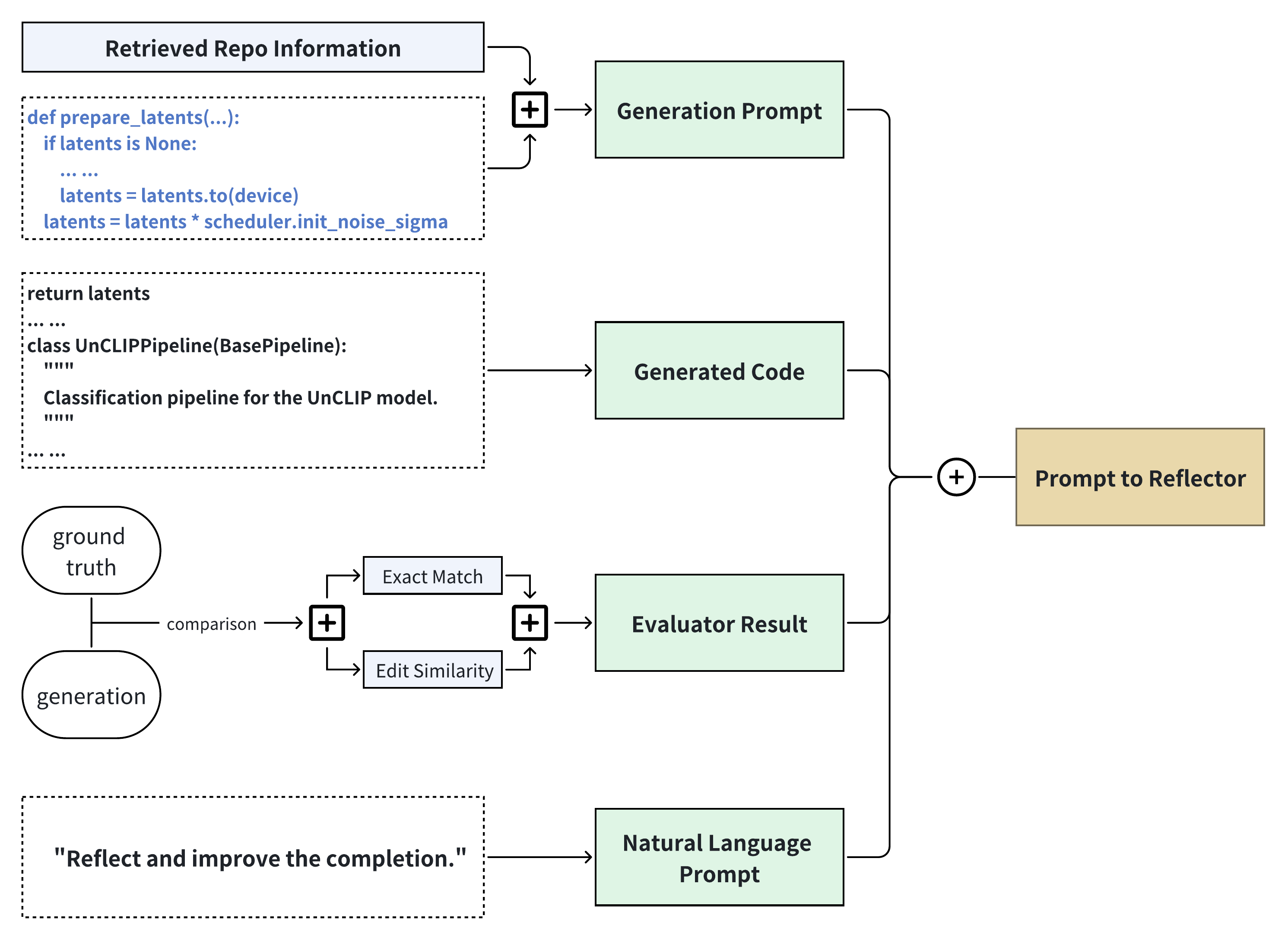}
    \caption{\small A process of the prompt concatenated to the Reflector component
}
    \label{fig:prompt2reflector}
\end{figure}

\subsubsection{Reflector}

Incorporating the Reflector component is a key innovation in our model design. 
The initial generation prompt, along with the Evaluator's results and the generated code, are concatenated to form a new prompt. 
We show the process of composing a prompt to Reflector in Figure~\ref{fig:prompt2reflector}.
The new prompt is fed back into the Reflector component to give detailed feedback. The Reflector allows the model to iteratively improve its outputs by learning from the evaluation results and its previous generation.
\subsubsection{Experience}
The Experience component stores the Reflector's feedback on each iteration. This historical data is used to inform the Retriever component, enhancing the relevance of future retrieved snippets based on the model's past performance. By leveraging past experiences, the model continuously refines its retrieval and generation processes.

\subsection{Leveraging Experience for Improved Retrieval}
\begin{figure}[h]
    \centering
    \includegraphics[width=0.48\textwidth]{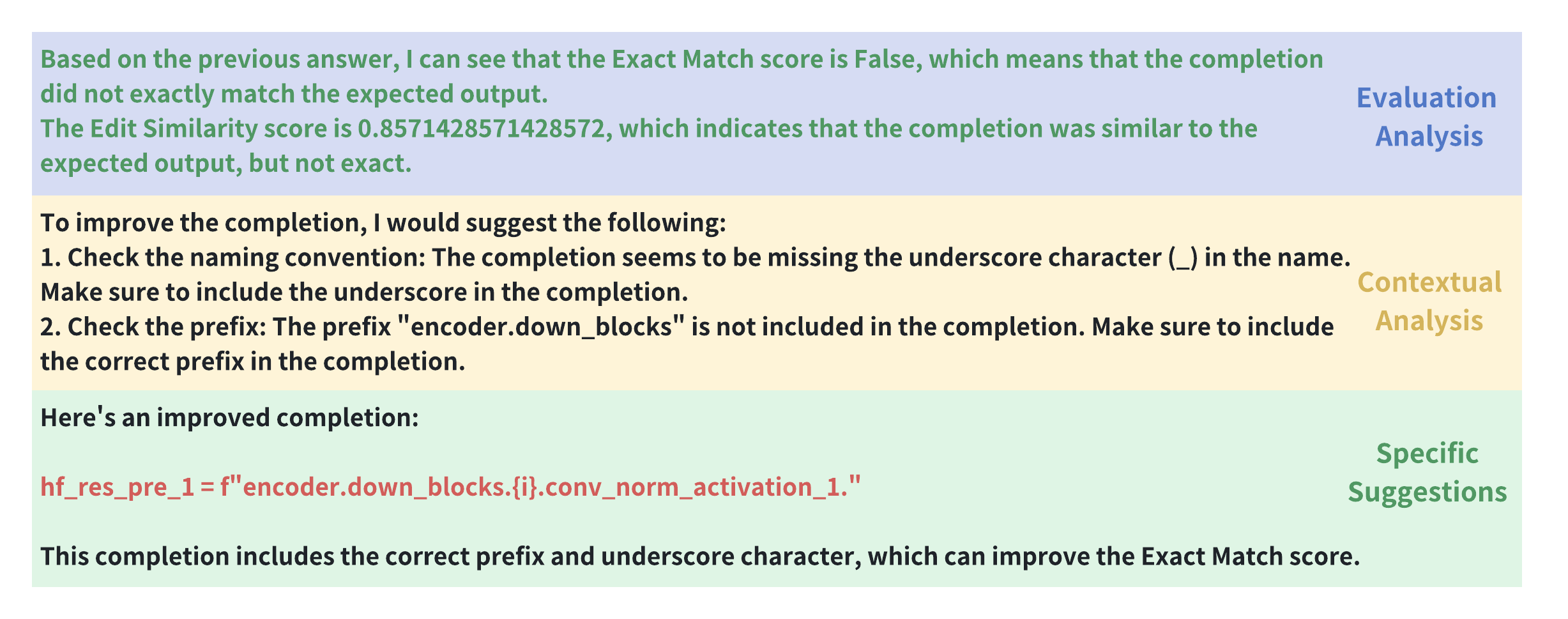}
    \caption{\small An example of detailed feedback from the Reflector, consisting of Evaluation Analysis, Contextual Analysis, and Specific Suggestions}
    \label{fig:reflector_feedback}
\end{figure}

The feedbacks stored in the Experience component play a crucial role in influencing the subsequent iterations. Each feedback generated by the Reflector component contains detailed analysis and suggestions to improve the generated code. As Figure \ref{fig:reflector_feedback} illustrates, this feedback includes:

\begin{itemize}
    \item \textbf{Evaluation Analysis}: An analysis of EM and ES scores that quantitatively assess the generated code.
    \item \textbf{Contextual Analysis}: An analysis of the syntax and semantics of the code, considering the broader context in which the code operates.
    \item \textbf{Specific Suggestions}: Recommendations for the generated code, including corrections and enhancements.
\end{itemize}

\begin{figure*}[h]
    \centering
    \includegraphics[width=\textwidth]{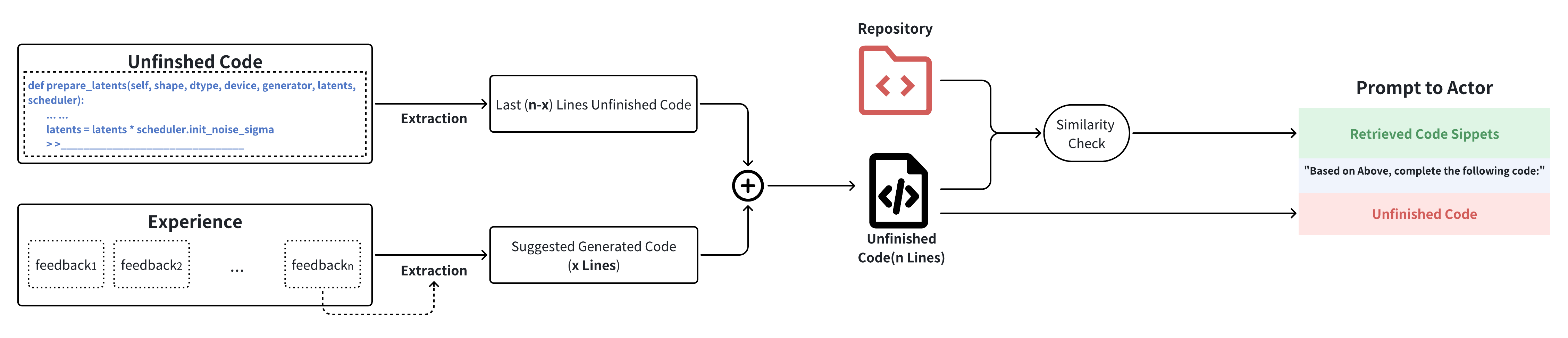}
    \caption{\small The process from Experience to next iteration}
    \label{fig:experience2retrieval}
\end{figure*}

These feedback elements are then extracted into the Reflector component for the next iteration. Figure \ref{fig:experience2retrieval} shows the details of the process.
Here’s how the procession works:
\begin{itemize}
    \item \textbf{Constructing New Retrieval Targets}: Each feedback provides a detailed analysis and suggested improvements. The suggested improvements from the feedback, occupying \(x\) lines, combined with the last \(n-x\) lines of the unfinished code, form a new \(n\)-line retrieval target. This can be mathematically represented as:

    \[
    T_{\text{new}} = \begin{bmatrix}
    F_{\text{feedback}} \\
    C_{\text{last}}
    \end{bmatrix}
    \]
    
    Here:
    
    \begin{itemize}
        \item \(F_{\text{feedback}}\) represents the suggested improvements in the feedback, which occupy \(x\) lines.
        \item \(C_{\text{last}}\) represents the remaining unfinished code, which occupies \(n-x\) lines.
    \end{itemize}

    \item \textbf{Similarity-Based Retrieval}: The new retrieval target is then used to perform a similarity search against the repository files. This process involves finding the most relevant code snippets that align with the retrieval target. By incorporating suggested improvements into the retrieval target, Retriever can better understand the context and the specific requirements of the code completion task. The similarity between the retrieval target and the repository snippets is quantified using the Jaccard Index, calculated as:
\[
\operatorname{Jaccard}(S_q, S_c) = \frac{|S_q \cap S_c|}{|S_q \cup S_c|}
\]

where \(S_q\) represents the set of tokens in the retrieval target, and \(S_c\) represents the set of tokens in a candidate code snippet. This index measures the similarity between the two sets, ensuring that the retrieval process identifies snippets with the most relevant shared elements.
\end{itemize}

This feedback-driven approach allows the framework to adapt and improve dynamically, leveraging historical performance data to guide future retrievals and generate more accurate and contextually appropriate code. By using detailed feedback to inform the retrieval targets, the framework can better align its generation process with the specific requirements of each task, ultimately leading to superior performance in code completion.

\subsection{Iterative Process}
In this section, we explore the iterative loop of our framework, focusing on how the Retriever, Actor, Evaluator, Reflector, and Experience components work together to refine generated code over multiple iterations. 

\subsubsection{Overall Iterative Process}

The model operates in a tightly coupled iterative loop, as illustrated in Figure \ref{fig:model-design}. The process starts with the Retriever extracting relevant snippets to generate a new prompt. The Actor uses this prompt to generate code, which is then evaluated. The Evaluator's results and generated code inform the Reflector, which refines the output and feeds it back to the Retriever, repeating the cycle.

\subsubsection{Achieving the Final Best Result in Iterations}

To determine the optimal stopping point in this process, we implement a stopping criterion based on the stability of evaluation scores. Specifically, the process stops if the ES score does not show significant improvement (e.g., less than 1\%) for a predefined number of iterations (e.g., 3 iterations) while the EM score is always 0. Algorithm \ref{alg:iterprocess} shows the detail.

\subsection{Component Interdependency}
The interdependency between the Reflector and Experience component is particularly important in driving the iterative process towards optimal results.

The Reflector's detailed feedback is the core input for the Experience component, which in turn enhances the retrieval process in subsequent iterations. This synergy between Reflector and Experience components drives the iterative process toward consistently higher quality code completions, ensuring robustness and adaptability in our framework.

\begin{algorithm}[t]
\caption{Iterative process for code completion}
\label{alg:iterprocess}
\begin{algorithmic}[1]
\Function{IterativeProcess}{$init\_input$}
\State Initialize $cur\_input \gets init\_input$, $iter\_cnt \gets 0$, $max\_iter \gets 10$, $best\_em \gets 0$, $best\_es \gets 0$
\State Initialize $no\_imp\_thres \gets 3$, $no\_imp\_cnt \gets 0$

\While{$iter\_cnt < max\_iter$}
    \State $retrieved\_snips \gets \text{Retriever}(cur\_input)$
    \State $gen\_code \gets \text{Actor}(retrieved\_snips)$
    \State $em, es \gets \text{Evaluator}(gen\_code)$
    \If{$em == 1.0$}
        \State \textbf{break} 
        \Comment{Correct result, stop iteration}
    \EndIf
    \If{$es - best\_es < 0.01$}
        \State $no\_imp\_cnt \gets no\_imp\_cnt + 1$
    \Else
        \State $no\_imp\_cnt \gets 0$
        \State $best\_em \gets em$, $best\_es \gets es$
    \EndIf
    \If{$no\_imp\_cnt \ge no\_imp\_thres$}
        \State \textbf{break} \Comment{Hardly improvement, stop iteration}
    \EndIf
    \State $feedback \gets \text{Reflector}(gen\_code, em, es)$
    \State $cur\_input \gets \text{UpdIptFdbk}(init\_input, fdbk)$
    \State $iter\_cnt \gets iter\_cnt + 1$
\EndWhile
\State \Return $gen\_code$
\EndFunction

\end{algorithmic}
\end{algorithm}


\section{Experiments}
\subsection{Benchmarks}
\label{data:benchmark}
While RepoEval~\cite{zhang2023repocoder} has been a foundational benchmark for assessing code completion across various levels of code granularity, including line, API invocation, and function body completion, its limitations hinder its ability to fully evaluate the complexities of modern, large-scale codebases. 
Since it includes repositories with a relatively low threshold for community engagement and a broad range of quality standards (e.g., 75\% of the repositories for line completion in RepoEval are more than 2 years old), it may not adequately reflect the real-world complexities and nuances faced by developers. 

To overcome these limitations, we introduce RepoGenEval, a new benchmark with stricter selection criteria focusing on high-quality repositories. By significantly raising the bar for community engagement --- requiring repositories to have between 700 to \num{40000} stars --- and ensuring that over 90\% of the code in each repository is Python with explicit unit tests, RepoGenEval better reflects the challenges of modern codebases. This benchmark is particularly tailored for line completion tasks, making it more relevant to practical applications of code completion.

Through these enhancements, RepoGenEval offers a robust, comprehensive platform for validating the effectiveness of our proposed framework.
%
\subsubsection{Construction of RepoGenEval}
\begin{table}[t]
\centering
\resizebox{0.48\textwidth}{!}{
\begin{tabular}{lcrrr}
\toprule
\textbf{Repository} & \textbf{Created} & \multicolumn{1}{c}{\textbf{Stars}} & \multicolumn{1}{c}{\textbf{Files}} & \multicolumn{1}{c}{\textbf{Lines}} \\
\midrule
\cite{MetaGPT}& 2023-06-30 & \num{40338} & 659 & \num{49659} \\
\cite{gpt-pilot}& 2023-08-16 & \num{ 28735}& \num{118} & \num{10686} \\
\cite{TaskWeaver} & 2023-09-11 & \num{4830} & 158 & \num{14905} \\
\cite{Otter} & 2023-04-01 & \num{3478} & 176 & \num{27967} \\
\cite{AutoGPT-ZH} & 2023-04-18 & \num{2381} & 143 & \num{10191} \\
\cite{OS-Copilot}  & 2024-02-02 & \num{1326} & 75 & \num{7770} \\
\cite{center-randomize}& 2024-04-05 & 748 & 8 & \num{736} \\
\cite{microagents} & 2023-12-11 & 706 & 45 & \num{2442} \\
\bottomrule
\end{tabular}
}
\caption{\small The repositories utilized in our RepoGenEval benchmark. Files denotes the total number of Python source files, and Lines denotes the total number of non-empty Python code lines.}
\label{tab:repositories}
\end{table}

In the development of the new benchmark, we meticulously curated a selection of Python repositories from GitHub, adhering to a set of rigorously defined criteria aimed at enhancing the quality and relevance of the benchmark for programming model validation. Each repository included in the benchmark is required to be an original, non-forked project with an open-source license, ensuring accessibility and encouraging community engagement. A significant enhancement in our selection criteria was the establishment of a higher threshold for community interaction, with a minimum requirement of 700 stars, extending up to \num{40000} stars (a significant increase from the 100-star minimum in the previous benchmark). This criterion ensures the inclusion of repositories that are not only widely recognized within the developer community but also demonstrate substantial and ongoing developer engagement, as evidenced by active issue interactions.

Moreover, to emphasize the technical quality and maintainability of the selected projects, we require that over 90\% of the content in each repository be Python code, accompanied by explicit unit tests. These criteria guarantee that the benchmark is highly relevant to the Python programming community and focuses on high-quality, maintainable code. To mitigate potential selection biases and promote a diverse representation of repository characteristics, we included the most recent, relevant projects, ensuring the dataset incorporates the latest coding standards and software technologies. We randomly selected repositories that satisfied the above criteria and that were created no earlier than 2023.

As shown in Table~\ref{tab:repositories}, the new benchmark provides a robust, comprehensive benchmark, comprising 8 high-quality repositories that reflect diverse applications and complexities. This improved benchmark is instrumental in validating the enhanced capabilities of our programming models.

We utilized GitHub's REST API and manual keyword searches to collect data on Python projects. From each repository, 200 lines of code were randomly selected for completion tasks, ensuring that these lines were unique, and not merely comments. This process resulted in a total of 1600 test samples for evaluating code completion.

\subsection{Selecting Optimal Reflector Model: Code Generation Models or General-purpose Models}

In our framework, the Actor and Reflector components are both large language models (LLMs). The Actor is responsible for completing unfinished code. The Reflector's role is to analyze the generated code and provide feedback to improve subsequent iterations. 
Given that the Actor focuses on generating code, we selected CodeGen-Mono-6B model due to its specialized design for program synthesis tasks. This model is particularly well-suited for handling complex code generation scenarios, offering robust performance and high accuracy in generating unfinished code, making it an ideal choice for our framework.

The Reflector primarily analyzes the code generated by the Actor and the corresponding evaluation results. In the context of code generation tasks, specialized language models pre-trained on extensive code datasets, like CodeGen-Mono-6B ~\cite{nijkamp2022codegen}, are likely to excel in analyzing and understanding code. These models are specifically designed for code-related tasks and so are adept at providing relevant feedback for code generation improvements.

However, the Reflector's purpose is to provide detailed feedback that goes beyond the initial generation to enhance the overall quality of the generated code. This requires the ability to offer comprehensive and nuanced feedback, a strength of general-purpose generative models pre-trained on diverse datasets like Meta-Llama-3-8B ~\cite{llama3modelcard}. These models can leverage their broad training to give more detailed and contextually rich feedback.

In this experiment, we selected two representative models from these categories to serve as our Reflector component:
\begin{itemize}
    \item \textbf{CodeGen-Mono-6B}: A specialized program synthesis model optimized for code generation tasks.
    \item \textbf{Meta-Llama-3-8B}: A general generative model designed for various natural language processing tasks.
\end{itemize}

We integrated each model into the Reflector component and evaluated their performance on the RepoGenEval benchmark using EM and ES metrics to measure the accuracy and quality of the code completions.

\begin{table}[h]
    \centering
    \resizebox{0.45\textwidth}{!}{%
    \begin{tabular}{lcccc}
        \toprule
        \multirow{2}{*}{Repository} & \multicolumn{2}{c}{EM} & \multicolumn{2}{c}{ES} \\
        \cmidrule(r){2-3} \cmidrule(r){4-5}
         & Model A & Model B & Model A & Model B \\
        \midrule
        Auto-GPT-ZH & 0.280 & 0.325 & 0.438 & 0.483 \\
        center-randomize & 0.305 & 0.350 & 0.470 & 0.520 \\
        gpt-pilot & 0.244 & 0.294 & 0.431 & 0.481 \\
        MetaGPT & 0.260 & 0.306 & 0.445 & 0.490 \\
        microagents & 0.270 & 0.322 & 0.450 & 0.510 \\
        OS-Copilot & 0.227 & 0.267 & 0.407 & 0.457 \\
        Otter & 0.255 & 0.310 & 0.440 & 0.490 \\
        TaskWeaver & 0.235 & 0.270 & 0.418 & 0.465 \\
        \bottomrule
    \end{tabular}
    }
    \caption{\small Comparison of EM and ES values for different Reflector models: codegen-6B-mono (Model A) vs. Meta-Llama-3-8B (Model B) with a fixed codegen-6B-mono Actor on RepoGenEval}
    \label{tab:reflexion_optimize}
\end{table}

\subsubsection{Results}
As shown in Table \ref{tab:reflexion_optimize}, the Meta-Llama-3-8B model consistently provided better feedback, leading to higher EM and ES scores across various repositories. This indicates that the general-purpose Meta-Llama-3-8B model is better suited for the Reflector component, as it enhances the framework's ability to refine code completions iteratively.

Based on these findings, we conclude that while specialized code generation models are effective for the Actor component, a general-purpose model like Meta-Llama-3-8B is more appropriate for the Reflector component, as it offers more comprehensive feedback and ultimately leads to better performance in code completion tasks.

\subsection{Overall Model Evaluation}

Following our exploration of the optimal Reflector model for our framework, we proceeded to evaluate the performance of the framework on the code completion task. To comprehensively assess its effectiveness, we compared it against different state-of-the-art (SOTA) models using our custom benchmark, RepoGenEval, which consists of the latest and high-quality real-world repositories, as well as the RepoEval benchmark. We selected four models for comparison:
\begin{table*}[h]
    \centering
    \resizebox{\textwidth}{!}{%
    \begin{tabular}{lcccccccccccccccc}
        \toprule
        \multirow{2}{*}{Model} & \multicolumn{2}{c}{diffusers} & \multicolumn{2}{c}{nerfstudio} & \multicolumn{2}{c}{fortuna} & \multicolumn{2}{c}{evaluate} & \multicolumn{2}{c}{vizier} & \multicolumn{2}{c}{FedScope} & \multicolumn{2}{c}{rl} & \multicolumn{2}{c}{ACE} \\
        \cmidrule(r){2-3} \cmidrule(r){4-5} \cmidrule(r){6-7} \cmidrule(r){8-9} \cmidrule(r){10-11} \cmidrule(r){12-13} \cmidrule(r){14-15} \cmidrule(r){16-17}
         & EM & ES & EM & ES & EM & ES & EM & ES & EM & ES & EM & ES & EM & ES & EM & ES \\
        \midrule
        CodeT5+ 2B & 0.427 & 0.715 & 0.425 & 0.713 & 0.429 & 0.718 & 0.424 & 0.714 & 0.431 & 0.716 & 0.428 & 0.713 & 0.426 & 0.714 & 0.425 & 0.715 \\
        CodeLlama-7b-hf & 0.440 & 0.722 & 0.437 & 0.720 & 0.441 & 0.725 & 0.436 & 0.721 & 0.443 & 0.723 & 0.440 & 0.720 & 0.438 & 0.721 & 0.437 & 0.722 \\
        StarCoder & 0.451 & 0.732 & 0.448 & 0.730 & 0.452 & 0.735 & 0.447 & 0.731 & 0.453 & 0.733 & 0.450 & 0.730 & 0.448 & 0.731 & 0.447 & 0.732 \\
        CodeGemma & 0.463 & 0.746 & 0.461 & 0.744 & 0.465 & 0.748 & 0.460 & 0.743 & 0.467 & 0.745 & 0.464 & 0.742 & 0.462 & 0.744 & 0.461 & 0.743 \\
        RepoGenReflex & \textbf{0.480} & \textbf{0.754} & \textbf{0.478} & \textbf{0.753} & \textbf{0.481} & \textbf{0.755} & \textbf{0.477} & \textbf{0.752}& \textbf{0.479} & \textbf{0.754} & \textbf{0.476} & \textbf{0.751} & \textbf{0.478} & \textbf{0.752} & \textbf{0.478} & \textbf{0.752} \\
        \bottomrule
    \end{tabular}
    }
    \caption{\small Comparison of EM and ES values for RepoGenReflex, CodeGemma, StarCoder, CodeLlama-7b-hf, and CodeT5+ 2B models on RepoEval benchmark.}
    \label{repoeval_results}
    \vspace{-10pt}
\end{table*}

\begin{table*}[h]
    \centering
    \resizebox{\textwidth}{!}{%
    \begin{tabular}{lcccccccccccccccc}
        \toprule
        \multirow{2}{*}{Model} & \multicolumn{2}{c}{Auto-GPT-ZH} & \multicolumn{2}{c}{center-randomize} & \multicolumn{2}{c}{gpt-pilot} & \multicolumn{2}{c}{MetaGPT} & \multicolumn{2}{c}{microagents} & \multicolumn{2}{c}{OS-Copilot} & \multicolumn{2}{c}{Otter} & \multicolumn{2}{c}{TaskWeaver} \\
        \cmidrule(r){2-3} \cmidrule(r){4-5} \cmidrule(r){6-7} \cmidrule(r){8-9} \cmidrule(r){10-11} \cmidrule(r){12-13} \cmidrule(r){14-15} \cmidrule(r){16-17}
         & EM & ES & EM & ES & EM & ES & EM & ES & EM & ES & EM & ES & EM & ES & EM & ES \\
        \midrule
        CodeT5+ 2B & 0.421 & 0.691 & 0.405 & 0.627 & 0.385 & 0.693 & 0.395 & 0.700 & 0.421 & 0.692 & 0.386 & 0.692 & 0.395 & 0.696 & 0.405 & 0.689 \\
        CodeLlama-7b-hf & 0.412 & 0.711 & 0.420 & 0.678 & 0.397 & 0.708 & 0.417 & 0.721 & 0.430 & 0.705 & 0.401 & 0.713 & 0.414 & 0.718 & 0.418 & 0.717 \\
        StarCoder & 0.422 & 0.706 & 0.417 & 0.713 & 0.405 & 0.713 & 0.415 & 0.720 & 0.442 & 0.721 & 0.406 & 0.712 & 0.412 & 0.721 & 0.425 & 0.712 \\
        CodeGemma & 0.436 & 0.721 & 0.435 & 0.726 & 0.412 & 0.719 & 0.421 & 0.728 & 0.441 & 0.719 & 0.411 & 0.720 & 0.421 & 0.718 & 0.422 & 0.725 \\
        RepoGenReflex & \textbf{0.438} & \textbf{0.735} & \textbf{0.439} & \textbf{0.738} &\textbf{0.420}  &\textbf{0.733}  &\textbf{0.430}  & \textbf{0.740} &\textbf{0.455}  &\textbf{0.735}  &\textbf{0.421}  & \textbf{0.732} &\textbf{0.430}  &\textbf{0.736}  &\textbf{0.439}  & \textbf{0.735}\\
        \bottomrule
    \end{tabular}
    }
    \caption{\small Comparison of EM and ES values for RepoGenReflex, CodeGemma, StarCoder, CodeLlama-7b-hf, and CodeT5+ 2B models on RepoGenEval benchmark.}
    \label{repogeneval_results}
    \vspace{-10pt}
\end{table*}
\begin{itemize}
    \item \textbf{CodeT5+ 2B}: A model optimized for code-related tasks, leveraging extensive training on large codebases to enhance its code generation capabilities.
    \item \textbf{CodeLlama-7b-hf}: A variant of the Llama models designed specifically for code synthesis and understanding, making it suitable for code completion tasks.
    \item \textbf{StarCoder}: A large language model specifically designed for code, trained on data from GitHub covering over 80 programming languages. It has shown superior performance on popular coding benchmarks.
    \item \textbf{CodeGemma}: Particularly the 7B parameter variants, which have demonstrated excellent performance in code completion tasks, especially on benchmarks like HumanEval Infilling and multilingual coding benchmarks.
\end{itemize}

We compared the performance of our framework against these models using EM+ES metrics. Unlike the SOTA models, which only utilized the JSONL file of our benchmark, our framework leveraged all elements in the benchmark and the iterative loop.

\subsubsection{Results}
The results, as shown in Table~\ref{repoeval_results} and Table~\ref{repogeneval_results}, demonstrate that while all selected models perform well, the RepoGenReflex framework consistently achieves higher EM and ES scores across most repositories. Specifically, our framework outperforms specialized and general-purpose SOTA models, particularly in complex repositories where contextual understanding is crucial. This indicates that RepoGenReflex is highly effective in providing accurate and contextually relevant code completions.

These findings underscore the robustness and adaptability of the RepoGenReflex framework, affirming its superior practical performance in standard code completion tasks.

\subsection{Ablation Study}
To make sure each component in our framework work, we make an ablation study to figure out whether each componont make the framework effective.

For this ablation study, we still keep CodeGen-Mono-6B as the Actor within the \fullname framework. By integrating CodeGen-Mono-6B with our Reflector model, we aim to evaluate how well the combined approach performs in terms of EM and ES metrics. 

\subsubsection{Experimental Setup}

In this ablation study, we aimed to evaluate the individual contributions of the Reflector, Experience, and Evaluator components of our \fullname framework. We performed the following configurations to systematically analyze the impact of these components:

\begin{itemize}
    \item \textbf{Full Model}: The full \fullname framework, incorporating Reflector, Experience, and Evaluator components, with CodeGen-Mono-6B as the actor and Meta-Llama-3-8B as the Reflector model.
    \item \textbf{Disabling Reflector and Experience}: This model excludes both the Reflector and Experience, relying solely on the base actor model (CodeGen-Mono-6B).
    \item \textbf{Disabling Evaluator}: This model excludes the Evaluator component, utilizing CodeGen-Mono-6B as the Actor and Meta-Llama-3-8B as the Reflector model.

\end{itemize}

\begin{table}[h]
    \centering
    \resizebox{0.48\textwidth}{!}{%
    \begin{tabular}{lcccccc}
        \toprule
        \multirow{2}{*}{Repository} & \multicolumn{3}{c}{EM} & \multicolumn{3}{c}{ES} \\
        \cmidrule(r){2-4} \cmidrule(r){5-7}
         & Model A & Model B & Full Model & Model A & Model B & Full Model \\
        \midrule
        diffusers & 0.365 & 0.445 & 0.480 & 0.665 & 0.744 & 0.754 \\
        nerfstudio & 0.360 & 0.440 & 0.478 & 0.660 & 0.740 & 0.753 \\
        fortuna & 0.355 & 0.430 & 0.481 & 0.655 & 0.734 & 0.755 \\
        evaluate & 0.350 & 0.435 & 0.477 & 0.650 & 0.741 & 0.752 \\
        vizier & 0.345 & 0.425 & 0.479 & 0.645 & 0.732 & 0.754 \\
        FedScope & 0.340 & 0.405 & 0.476 & 0.640 & 0.721 & 0.751 \\
        rl & 0.355 & 0.435 & 0.478 & 0.655 & 0.740 & 0.752 \\
        ACE & 0.350 & 0.415 & 0.478 & 0.650 & 0.732 & 0.752 \\
        \bottomrule
    \end{tabular}
    }
    \caption{\small Comparison of EM and ES values for models across on RepoEval benchmark. Model A: without Reflector and Experience components, Model B: without Evaluator components, Full Model: baseline model.}
    \label{repoeval_ablation_results}
    \vspace{-10pt}
\end{table}

We used standard code completion benchmark RepoEval and assessed performance using EM and ES metrics. Table~\ref{repoeval_ablation_results} shows the results.

\subsubsection{Results}
The ablation study results highlight the critical roles of the Reflector, Experience, and Evaluator components in enhancing code generation quality. Disabling Reflector and Experience led to a significant decline in EM and ES scores, indicating their joint importance. The Evaluator component also showed a considerable impact on performance, underscoring its value in providing detailed feedback and guiding the iterative improvement process.

\section{Conclusion}
In this paper, we presented RepoGenReflex, a novel code completion framework that effectively integrates RAG with VRL to optimize the iterative retrieval and generation process. By leveraging a carefully selected combination of large language models for the Reflector components, RepoGenReflex proved to be outperforming SOTA models in higher accuracy for code completion tasks. We also validated the critical roles of the Reflector, Experience, and Evaluator components in enhancing code generation quality. These advancements provide a robust foundation for future research and development in intelligent code generation and completion. Our work demonstrates the potential of combining RAG with VRL to create a versatile and powerful code completion framework.

\bibliography{references}

\end{document}